\documentstyle[preprint,aps,prl]{revtex}
\begin{document}
\draft
\preprint{\parbox{2.2in}{FERMILAB--PUB--95/045--T\\[-12pt]CLNS 95/1329
\\[-12pt]hep-ph/9503356\\[12pt]}}
\title{Quarkonium Wave Functions at the Origin}
\author{Estia J. Eichten$^{\rm a}$\thanks{Internet address:
eichten@fnal.gov}
and Chris Quigg$^{\rm a,b}$\thanks{Internet address: quigg@fnal.gov}}
\address{$^{\rm a)}$Theoretical Physics Department\\
Fermi National Accelerator Laboratory\\
P.O. Box 500, Batavia, Illinois 60510 \\
and \\$^{\rm b)}$Floyd R. Newman Laboratory of Nuclear Studies \\
Cornell University \\ Ithaca, New York 14853}
\date{\today}
\maketitle
\begin{abstract}
We tabulate values of the radial Schr\"{o}dinger wave function or its
first nonvanishing derivative at zero quark-antiquark separation, for
$c\bar{c}$, $c\bar{b}$,  and $b\bar{b}$ levels that lie below, or
just above, flavor
threshold.  These quantities are essential inputs for evaluating
production cross sections for quarkonium states.
\end{abstract}
%\section{Introduction}
%\narrowtext
Fragmentation of partons into quarkonium states recently has been
recognized as an important component of quarkonium production in
high-energy collisions \cite{Braaten}.  Thorough investigation of this
new source, and others, is made timely by the CDF Collaboration's
observation that
conventional sources substantially underestimate the yield of prompt
$J\!/\!\psi$ and $\psi^\prime$ in 1.8-TeV $\bar{p}p$ collisions
\cite{psiprprob}.  In the case of the $\psi^\prime$, which is not fed by
the cascade decay of the narrow $\chi_c$ states, the observed cross
section exceeds the calculated one by more than an order of magnitude.

Calculation of the production rate by fragmentation can be separated into
a parton-level piece that can be evaluated using perturbative techniques
and a hadronic piece expressed in terms of the quarkonium wave function.
We have earlier tabulated the values at the origin of the radial wave
function, or its first nonvanishing derivative, for narrow levels of the
$b\bar{c}$ ($B_c$) system \cite{B_c}.  These have been used to estimate
the $B_c$ production rate \cite{B_cProd}.  Here we present the corresponding
information for the $c\bar{c}$ ($J\!/\!\psi$) and $b\bar{b}$ ($\Upsilon$)
families.  Although many of these numbers have appeared in the literature
when quarkonium spectroscopy was in flower, they usually were given
implicitly in calculations of leptonic widths or similar observables.
Our purpose here is to record the relevant properties of the wave
functions of all the narrow levels in a form convenient for evaluating
cross sections for quarkonium production.

We consider four functional forms for the potential that give reasonable
accounts of the $c\overline{c}$ and $b\overline{b}$ spectra: the
QCD-motivated potential given by Buchm\"{u}ller and
Tye \cite{BT}, with
\begin{eqnarray}
m_c = 1.48\text{ GeV}\!/\!c^2 & \;\;\; &m_b = 4.88\text{ GeV}\!/\!c^2\;\;;
\end{eqnarray}
a power-law potential \cite{Martin},
\begin{equation}
	V(r) = -8.064\text{ GeV} + (6.898\text{ GeV})(r\cdot\text{1 GeV})^{0.1}
	 \;\;, \label{Martin}
\end{equation}
with
\begin{eqnarray}
m_c = 1.8\text{ GeV}\!/\!c^2 & \;\;\; & m_b = 5.174\text{ GeV}\!/\!c^2\;\;;
\end{eqnarray}
a logarithmic potential \cite{Log},
\begin{equation}
	V(r) = -0.6635\text{ GeV} + (0.733\text{ GeV}) \log{(r\cdot\text{1
	GeV})}\;\;,
\end{equation}
with
\begin{eqnarray}
m_c = 1.5\text{ GeV}\!/\!c^2 & \;\;\; & m_b = 4.906\text{ GeV}\!/\!c^2\;\;;
\end{eqnarray}
and a Coulomb-plus-linear potential (the ``Cornell potential'')
\cite{Cornell},
\begin{equation}
	V(r)=-\frac{\kappa}{r} + \frac{r}{a^2}\;\;,
	\label{cornellpot}
\end{equation}
with
\begin{eqnarray}
m_c = 1.84\text{ GeV}\!/\!c^2 & \;\;\; & m_b = 5.18\text{ GeV}\!/\!c^2 \\
\kappa=0.52 & \;\;\; & a=2.34\text{ GeV}^{-1}\;\;.
\end{eqnarray}

For quarks bound in a central potential, it is convenient to separate
the Schr\"{o}dinger wave function into
radial and angular pieces as
$\Psi_{n\ell m}(\vec{r}) = R_{n\ell}(r)Y_{\ell m}(\theta,\phi)$,
where $n$ is the principal quantum number, $\ell$ and $m$ are the
orbital angular momentum and its projection, $R_{n\ell}(r)$ is the radial
wave function, and $Y_{\ell m}(\theta,\phi)$ is a spherical harmonic
\cite{ylm}.  The Schr\"{o}dinger wave function is normalized,
	$\int{ d^3\vec{r} |\Psi_{n\ell m}(\vec{r})|^2} = 1$,
so that
$\int_0^\infty r^2 dr |R_{n\ell}(r)| = 1$.
The value of the radial wave
function, or its first nonvanishing derivative at the origin,
\begin{equation}
	R_{n\ell}^{(\ell)}(0)\equiv \left.\frac{d^{\,\ell}R_{n\ell}(r)}
	{dr^\ell}\right|_{r=0}\;\;\;,
	\label{wvfc}
\end{equation}
is required to evaluate production rates
through parton fragmentation.  The quantity
$|R_{n\ell}^{(\ell)}(0)|^2$ is presented for four potentials in Table
\ref{psizerocc} for the narrow charmonium levels and in Table
\ref{psizerobb} for the narrow Upsilon levels.  For ease of reference, we
reproduce in Table \ref{psizerobc} our predictions \cite{B_c} for the
$B_c$ wave functions, with some computational improvements in the entries
for the Cornell potential.  The
strong Coulomb singularity of the Cornell potential is
reflected in spatially smaller states.

In view of the efforts \cite{above} to resolve the $\psi^\prime$ anomaly
with cascades from above-threshold states, we quote values for
the $c\bar{c}$ 3D, 3P, and 3S levels that lie near 3.8, 3.9, and
$4.0\text{ GeV}\!/\!c^2$, respectively, and for the $b\bar{b}$ 4F, 4D, 4P,
and 4S levels that lie near 10.35, 10.45, 10.52, and
$10.6\text{ GeV}\!/\!c^2$,
respectively.  It  is likely that these will be
significantly modified by coupled-channel effects \cite{Cornell,coupled}.

If all the potentials describe the $c\bar{c}$ and $b\bar{b}$ observables,
what accounts for the wide variation in the values of
$|R_{n0}(0)|^2$ and the corresponding quantities for orbital
excitations?  The leptonic decay rate of a neutral
($Q\bar{Q}$) vector meson $V^0$ is
related to the Schr\"{o}dinger wave function through \cite{vrw,QCDRC}
\begin{equation}
	\Gamma(V^0\rightarrow e^+ e^-) = \frac{4 N_c \alpha^2 e_Q^2}{3}
	\frac{|R_{n0}(0)|^2}{M_V^2} \left(1-{\displaystyle
	\frac{16\alpha_s}{3\pi}}\right)\;\;,
	\label{vrwf}
\end{equation}
where $N_c=3$ is the number of quark colors, $e_Q$ is the heavy-quark
charge, and $M_V$ is the mass of the vector meson.  The QCD correction
reduces  the magnitudes
significantly; the amount of this reduction is somewhat uncertain,
because the first term in the perturbation expansion is large
\cite{correct}.  Each of the potentials we use corresponds to a
particular interpretation of the radiative correction to the leptonic
width.  Similar effects may enter the connection between wave functions
and fragmentation probabilities.

\acknowledgments
We thank Peter Cho for posing the question that led to this note, and we
thank Eric Braaten for helpful comments on the manuscript.

Fermilab is operated by Universities Research Association, Inc., under
contract DE-AC02-76CHO3000 with the United States Department of Energy.
CQ thanks the Department of Physics and Laboratory of Nuclear Studies at
Cornell University for warm hospitality.

\widetext
\begin{table}
\caption{Radial wave functions at the origin and related quantities for
$c\bar{c}$ mesons.}
\begin{tabular}{ccccc}
Level & \multicolumn{3}{c}{$|R_{n\ell}^{(\ell)}(0)|^2$} \\
 & QCD (B--T), Ref. \cite{BT} & Power-law, Ref. \cite{Martin} & Logarithmic,
 Ref. \cite{Log} & Cornell, Ref. \cite{Cornell}  \\
\tableline
1S & $0.810\text{ GeV}^3$ & $0.999\text{ GeV}^3$ & $0.815\text{ GeV}^3$ &
$1.454\text{ GeV}^3$ \\
2P & $0.075\text{ GeV}^5$ & $0.125\text{ GeV}^5$ & $0.078\text{ GeV}^5$ &
$0.131\text{ GeV}^5$ \\
2S & $0.529\text{ GeV}^3$ & $0.559\text{ GeV}^3$ & $0.418\text{ GeV}^3$ &
$0.927\text{ GeV}^3$ \\
\hline
3D & $0.015\text{ GeV}^7$ & $0.026\text{ GeV}^7$ & $0.012\text{ GeV}^7$ &
$0.031\text{ GeV}^7$ \\
3P & $0.102\text{ GeV}^5$ & $0.131\text{ GeV}^5$ & $0.076\text{ GeV}^5$ &
$0.186\text{ GeV}^5$ \\
3S & $0.455\text{ GeV}^3$ & $0.410\text{ GeV}^3$ & $0.286\text{ GeV}^3$ &
$0.791\text{ GeV}^3$
\end{tabular}
\label{psizerocc}
\end{table}

\begin{table}
\caption{Radial wave functions at the origin and related quantities for
$b\bar{b}$ mesons.}
\begin{tabular}{ccccc}
Level & \multicolumn{3}{c}{$|R_{n\ell}^{(\ell)}(0)|^2$} \\
 & QCD (B--T), Ref. \cite{BT} & Power-law, Ref. \cite{Martin} & Logarithmic,
 Ref. \cite{Log} & Cornell, Ref. \cite{Cornell}  \\
\tableline
1S & $6.477\text{ GeV}^3$ & $4.591\text{ GeV}^3$ & $4.916\text{ GeV}^3$ &
$14.05\text{ GeV}^3$ \\
2P & $1.417\text{ GeV}^5$ & $1.572\text{ GeV}^5$ & $1.535\text{ GeV}^5$ &
$2.067\text{ GeV}^5$ \\
2S & $3.234\text{ GeV}^3$ & $2.571\text{ GeV}^3$ & $2.532\text{ GeV}^3$ &
$5.668\text{ GeV}^3$ \\
3D & $0.637\text{ GeV}^7$ & $0.892\text{ GeV}^7$ & $0.765\text{ GeV}^7$ &
$0.860\text{ GeV}^7$ \\
3P & $1.653\text{ GeV}^5$ & $1.660\text{ GeV}^5$ & $1.513\text{ GeV}^5$ &
$2.440\text{ GeV}^5$ \\
3S & $2.474\text{ GeV}^3$ & $1.858\text{ GeV}^3$ & $1.736\text{ GeV}^3$ &
$4.271\text{ GeV}^3$ \\
4F & $0.414\text{ GeV}^9$ & $0.627\text{ GeV}^9$ & $0.456\text{ GeV}^9$ &
$0.563\text{ GeV}^9$ \\
4D & $1.191\text{ GeV}^7$ & $1.396\text{ GeV}^7$ & $1.119\text{ GeV}^7$ &
$1.636\text{ GeV}^7$ \\
4P & $1.794\text{ GeV}^5$ & $1.593\text{ GeV}^5$ & $1.377\text{ GeV}^5$ &
$2.700\text{ GeV}^5$ \\
\hline
4S & $2.146\text{ GeV}^3$ & $1.471\text{ GeV}^3$ & $1.324\text{ GeV}^3$ &
$3.663\text{ GeV}^3$ \\
5F & $1.075\text{ GeV}^9$ & $1.302\text{ GeV}^9$ & $0.894\text{ GeV}^9$ &
$1.520\text{ GeV}^9$ \\
5D & $1.722\text{ GeV}^7$ & $1.689\text{ GeV}^7$ & $1.289\text{ GeV}^7$ &
$2.417\text{ GeV}^7$ \\
5P & $1.935\text{ GeV}^5$ & $1.504\text{ GeV}^5$ & $1.252\text{ GeV}^5$ &
$2.917\text{ GeV}^5$ \\
5S & $1.956\text{ GeV}^3$ & $1.231\text{ GeV}^3$ & $1.077\text{ GeV}^3$ &
$3.319\text{ GeV}^3$

\end{tabular}
\label{psizerobb}
\end{table}
\begin{table}
\caption{Radial wave functions at the origin and related quantities for
$c\bar{b}$ mesons.}
\begin{tabular}{ccccc}
Level & \multicolumn{3}{c}{$|R_{n\ell}^{(\ell)}(0)|^2$} \\
 & QCD (B--T), Ref. \cite{BT} & Power-law, Ref. \cite{Martin} & Logarithmic,
 Ref. \cite{Log} & Cornell, Ref. \cite{Cornell}  \\
\tableline
1S & $1.642\text{ GeV}^3$ & $1.710\text{ GeV}^3$ & $1.508\text{ GeV}^3$ &
$3.184\text{ GeV}^3$ \\
2P & $0.201\text{ GeV}^5$ & $0.327\text{ GeV}^5$ & $0.239\text{ GeV}^5$ &
$0.342\text{ GeV}^5$ \\
2S & $0.983\text{ GeV}^3$ & $0.950\text{ GeV}^3$ & $0.770\text{ GeV}^3$ &
$1.764\text{ GeV}^3$ \\
3D & $0.055\text{ GeV}^7$ & $0.101\text{ GeV}^7$ & $0.055\text{ GeV}^7$ &
$0.102\text{ GeV}^7$ \\
\hline
3P & $0.264\text{ GeV}^5$ & $0.352\text{ GeV}^5$ & $0.239\text{ GeV}^5$ &
$0.461\text{ GeV}^5$ \\
3S & $0.817\text{ GeV}^3$ & $0.680\text{ GeV}^3$ & $0.563\text{ GeV}^3$ &
$1.444\text{ GeV}^3$
\end{tabular}
\label{psizerobc}
\end{table}


\begin{references}
\bibitem{correct} If, for example, we interpret the factor
$(1-16\alpha_s/3\pi)$ as the beginning of an expansion for
$(1+16\alpha_s/3\pi)^{-1}$ with $\alpha_s=0.36$, then the
Buchm\"{u}ller--Tye-potential predictions for
the $\psi$ family agree with experiment, within errors, while those for the
$\Upsilon$ family are about $20\%$ low.
\end{references}
\end{document}